\documentstyle[aps,pre,epsf,rotate,amsfonts]{revtex}

\newcommand{\figurewidth}{10cm}
\begin{document}

\draft

\title{Optimized energy calculation in lattice systems with long-range
       interactions}

\author{Michael Krech}
\address{Institut f\"ur Theoretische Physik, RWTH Aachen,
         D-52056 Aachen, Germany}

\author{Erik Luijten}
\address{Max-Planck-Institut f\"ur Polymerforschung, Postfach 3148,
         D-55021 Mainz, Germany}
\address{Institut f\"ur Physik, WA 331, Johannes Gutenberg-Universit\"at,
         D-55099 Mainz, Germany}

\date{\today}

\maketitle

\begin{abstract}
  We discuss an efficient approach to the calculation of the internal energy in
  numerical simulations of spin systems with long-range interactions. Although,
  since the introduction of the Luijten--Bl\"ote algorithm, Monte Carlo
  simulations of these systems no longer pose a fundamental problem, the energy
  calculation is still an ${\cal O}(N^2)$ problem for systems of size~$N$.
  We show how this can be reduced to an ${\cal O}(N \log N)$ problem,
  with a break-even point that is already reached for very small systems. This
  allows the study of a variety of, until now hardly accessible, physical
  aspects of these systems. In particular, we combine the optimized energy
  calculation with histogram interpolation methods to investigate the specific
  heat of the Ising model and the first-order regime of the three-state Potts
  model with long-range interactions.
\end{abstract}


\section{Introduction}
The numerical study of systems with long-range interactions is notoriously
difficult, due to the large number of interactions that has to be taken into
account. Specifically, the number of operations to calculate the energy of a
single particle scales as the total number of particles in the system, in
contrast to the case of short-range interactions, where the corresponding
number of operations is of order unity. This implies that Monte Carlo-based
methods are generally restricted to very small system sizes, which are still
hampered by strong finite-size effects. Some years ago, this problem was
resolved for the case of ${\rm O}(n)$ spin systems with (ferromagnetic)
long-range interactions, for which a dedicated cluster algorithm was
developed~\cite{lr-alg}.  Since the efficiency of this algorithm is {\em
independent\/} of the number of interactions per spin, speed improvements of
several orders of magnitude could be obtained compared to a conventional
cluster algorithm. This speed-up pertains to the generation of independent
configurations, for which the calculation of the energy is not required.
Indeed, a variety of interesting physical results could be obtained by means of
this method, see, {\em e.g.}, Refs.~\cite{renorm,chicross,mr3d}. Whenever one
needs to sample the internal energy, however, the improvement is much less
dramatic: the major remaining advantage is that one only has to calculate the
energy for truly independent configurations, rather than in every Monte Carlo
step.  Whereas this still implies that one can study systems which are an order
of magnitude larger than those that can be accessed via Metropolis-type
simulations (cf.\ Ref.~\cite{glumac98}), one is eventually limited by the fact
that the total computing time scales quadratically with the system size.  One
major disadvantage of this inaccessibility of the energy is the fact that it is
not possible to apply histogram interpolations in order to obtain information
on thermodynamic quantities over a large parameter space~\cite{multiple-hist}.
In this paper, we point out that, for systems with periodic boundary
conditions, this problem can be circumvented by calculating the internal energy
in momentum space. Thus, one can apply a Fast Fourier Transform~(FFT), reducing
the total computational effort to ${\cal O}(N \log N)$ for a system containing
$N$ spins.  Indeed, this is a natural choice if one recognizes that the total
energy is just given by a (discrete) convolution, which is one of the major
applications of the FFT\@.  Remarkably, the computational overhead entailed by
the FFT turns out not to be a limiting factor: already for very small systems
it is more efficient than a direct calculation of the energy.

The remainder of this paper is organized as follows. First, we derive an
expression for the energy in terms of the Fourier-transformed spin system and
point out that several other observables can be obtained on the fly, at
negligible additional cost. We also give a detailed comparison of our approach
and the conventional method.  Next, we illustrate our approach by means of
several new physical results for one-dimensional systems with long-range
interactions.  We end with a summary of our results.
 
\section{Energy calculation}

We will now first illustrate our approach for a $(d=1)$-dimensional system with
an $n$-component order parameter, i.e., a generalized ${\rm O}(n)$ spin chain.
This system is described by the Hamiltonian
\begin{equation}
\label{eq:1dhamil}
  {\cal H} = -\frac{1}{2} \sum_{x=1}^{N} \sum_{y=1}^{N} J(x-y)
  {\bf S}(x) \cdot {\bf S}(y) \;,
\end{equation}
where the spins ${\bf S}(x)$ are $n$-component unit vectors and $N$ is the
system size. The interaction $J(x)$ is defined for all $x \in \Bbb{N}$.  Under
the condition that periodic boundary conditions are employed, the {\em
effective\/} coupling $\tilde{J}(x)$ between two spins is given by the sum over
all periodic copies,
\begin{equation}
\label{eq:Jofx}
\tilde{J}(x) \equiv \sum_{m=-\infty}^{\infty} J(x+mN)
\end{equation}
and hence has a period~$N$. We set the self energy~$\tilde{J}(mN)$, which is
just an additive constant in the total energy, equal to zero.  Each component
of the spin configuration ${\bf S}(x)$ and the interaction $\tilde{J}(x)$ can
then be written as a Fourier sum
\begin{equation}
\label{eq:s-ft}
  f(x) = \frac{1}{\sqrt{N}} \sum_{k=0}^{N-1} f_k e^{2\pi i k x/N},
\end{equation}
where the Fourier coefficients~$f_k$ are obtained from the discrete Fourier
transform of $f(x)$. By means of the discrete convolution
theorem~\cite{brigham} it is then straightforward to show that
Eq.~(\ref{eq:1dhamil}) can be written as
\begin{equation}
\label{eq:1dhamil-ft}
  {\cal H} = -\frac{\sqrt{N}}{2} \sum_{k=0}^{N-1} \tilde{J}_k
  {\bf S}_k \cdot {\bf S}_{-k} \;.
\end{equation}
The essential step is now, that application of the Fast Fourier
Transform~\cite{brigham,cooley65} reduces the computational effort for the
calculation of the $N$ Fourier coefficients from ${\cal O}(N^2)$ to ${\cal O}
(N \log N)$, thus, in principle, greatly speeding up the calculation of the
total energy.  The sum in~(\ref{eq:1dhamil-ft}) adds another ${\cal O}(N)$
operations, but this is compensated for by the fact that one typically also
wants to calculate the magnetic susceptibility $N^{-1}|\sum_{x=0}^{N-1}
{\bf S}(x)|^2$, which in the momentum-space representation is immediately
given by $|{\bf S}_{k=0}|^2$. For maximum efficiency, one has to restrict the
system size to powers of~2. Naturally, the calculation of the coefficients
$\tilde{J}_k$ has to be carried out only once.

Even more can be gained, if one also desires to calculate the spin--spin
correlation function $g(r) = \langle {\bf S}(0) \cdot {\bf S}(r) \rangle =
N^{-1} \sum_{x=0}^{N-1} {\bf S}(x) \cdot {\bf S}(x+r)$. The discrete
correlation theorem~\cite{brigham} states that the Fourier transform $g_k$ of
$g(r)$ is equal to $N^{-1/2} {\bf S}_k \cdot {\bf S}_{-k}$, so that $g_k$
is obtained by $N$ multiplications rather than another ${\cal O}(N^2)$
operations in the real-space representation.

\begin{figure}
\centering\leavevmode
\epsfxsize \figurewidth
\epsfbox{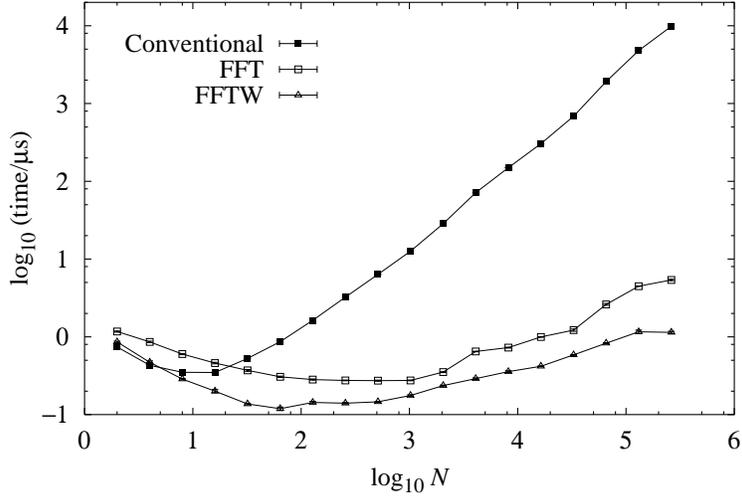}
\caption[]{Required CPU time (in microseconds on a Pentium-II 400 MHz) {\em per
spin\/} for the calculation of the energy and the susceptibility, as a function
of system size, for both the conventional method and two FFT-based methods. For
a further discussion see the text.}
\label{fig:timing}
\end{figure}

All the above estimates are only measures for the complexity of the algorithm,
which become valid for sufficiently large~$N$. It remains to be seen whether
the FFT-based approach is actually faster for the range of system sizes that
can be accessed in present-day Monte Carlo simulations, which for lattice
models go up to $N \sim 10^5-10^6$. Figure~\ref{fig:timing} compares the
required CPU time per spin for the calculation of the internal energy via
Eq.~(\ref{eq:1dhamil}) and Eq.~(\ref{eq:1dhamil-ft}), respectively, and the
susceptibility. As expected, the former method scales asymptotically linearly
with~$N$. For the latter method, two estimates are given, which only differ in
the choice of the FFT implementation. The slower results (open squares) were
obtained by means of the routines of Ref.~\cite{nr} and the faster
(triangles) by means of those of Ref.~\cite{fftw}. Although these two estimates
differ by as much as a factor of~$5$, both of them outperform the conventional
method already for $N \gtrsim 10$; for $N=2^{18}$ the improvement amounts to
roughly four orders of magnitude. The initial downward trend in
Fig.~\ref{fig:timing} is due to overhead being distributed over an increasing
number of spins. Likewise can the ``irregularities'' in the FFT estimates be
attributed to computational aspects. The slight deviations from linearity in
the conventional estimates, however, are due to statistical inaccuracies in the
timing: for $N \gtrsim {\cal O}(10^4)$ this method becomes already
prohibitively slow. Thus, we conclude that the method presented here provides a
very efficient approach to energy calculations in lattice systems with
long-range interactions; while there is still a weak system-size dependence
in the required computational effort per spin, this no longer constitutes a
bottleneck for practical applications. Note that higher-dimensional models
can also be treated in this fashion.

\section{Applications}
\subsection{The Ising chain}
As a first example, we consider the Ising chain with algebraically decaying
interactions, $J(x) \equiv J |x|^{-1-\sigma}$.  The critical behavior for this
system is essentially classical for $0 < \sigma \leq \frac{1}{2}$ and
nonclassical for $\frac{1}{2} < \sigma \leq 1$, see Ref.~\cite{fisher72b}.
Numerical results for the thermal exponent~$y_t$~\cite{thesis} have indicated
that the latter regime can be subdivided into two parts: $y_t > \frac{1}{2}$
for $\frac{1}{2} < \sigma \lesssim 0.65$ and $y_t < \frac{1}{2}$ for
interactions that decay faster. This implies that the specific heat only
diverges in a part of the nonclassical regime and should display a cusp-like
singularity in the remaining part. By means of illustration, we have calculated
the specific heat for $\sigma=0.25$ and $\sigma=0.90$. In both cases, we expect
to find a function that does not diverge at the critical point, although the
behavior should be qualitatively different. Simulations were carried out for
$N=2^p$, with $3 \leq p \leq 16$, at a number of different couplings, for
several times~$10^6$ independent samples per system size.  The full curves were
determined by means of the multiple-histogram method~\cite{multiple-hist},
where great care was taken to minimize systematic errors due to the histogram
interpolation.

\begin{figure}
\centering\leavevmode
\epsfxsize \figurewidth
\epsfbox{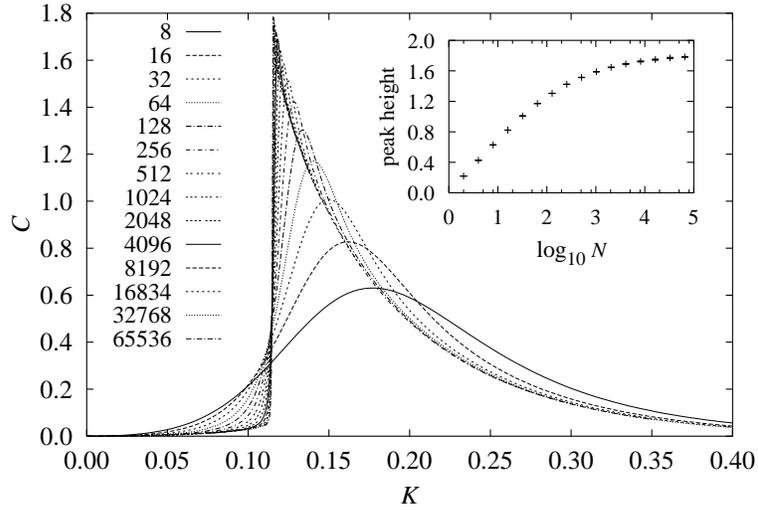}
\caption[]{Specific heat of the Ising chain with interactions decaying as
$r^{-1.25}$, for system sizes $8 < N < 65536$. One observes the appearance of a
mean-field-like discontinuity at the critical coupling. The inset shows the
peak height as a function of system size.}
\label{fig:c_125}
\end{figure}

Figure~\ref{fig:c_125} shows the specific heat~$C$ for $\sigma=0.25$, as a
function of the reduced coupling~$K \equiv J/(k_{\rm B}T)$. It displays
several close similarities to the specific heat of the mean-field model,
including the build-up of a jump discontinuity at the critical point, the
crossing of the finite-size curves in a single point at $K_c$ (up to
corrections to scaling), and (not visible on this scale) an excess peak in the
curves for finite systems, i.e., $\lim_{N \to \infty} C_{\rm max}(N) \neq
\lim_{K \downarrow K_c} \lim_{N \to \infty} C(K,N)$~\cite{thesis}.  As shown in
Ref.~\cite{renorm}, the location of the specific-heat maximum shifts as a
function of system size, according to
\begin{equation}
 K_{\rm max} = K_c + a_1 L^{-y_t^*} + a_2 L^{-2y_t^*} + b_1 L^{\sigma-1} +
 \cdots \;,
\end{equation}
where $y_t^* = \frac{1}{2}$ and the coefficients $a_i, b_i$ are nonuniversal.
A fit to this expression yielded $y_t^* = 0.51~(6)$ and $K_c = 0.1147~(5)$, in
good agreement with $K_c = 0.114142~(2)$~\cite{class-lr}. The inset shows the
peak height as a function of system size, strongly suggesting that the maximum
is indeed finite in the thermodynamic limit.

\begin{figure}
\centering\leavevmode
\epsfxsize \figurewidth
\epsfbox{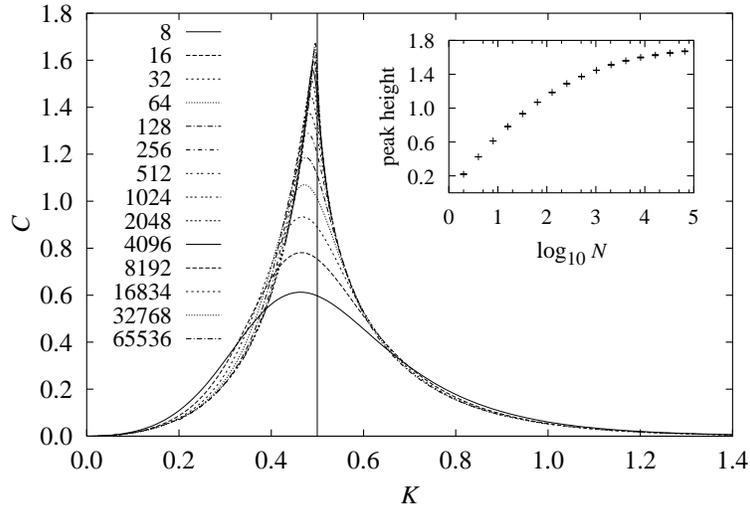}
\caption[]{Specific heat of the Ising chain with interactions decaying as
$r^{-1.90}$, for system sizes $8 < N < 65536$. With increasing~$N$, a cusp-like
singularity appears.  The vertical line indicates the critical coupling. The
inset shows the peak height as a function of system size.}
\label{fig:c_190}
\end{figure}

The case $\sigma=0.90$, shown in Fig.~\ref{fig:c_190}, clearly exhibits a
distinctly different behavior. The specific heat is now nonzero in the
thermodynamic limit, on either side of the critical point, and indeed displays
the expected cusp-like singularity. The inset confirms that the maximum is
convergent for $N \to \infty$. Since $y_t$ is still sufficiently close to
$\frac{1}{2}$, i.e., the absolute value of the exponent $\alpha$ is
sufficiently small, the location of the maximum cannot be distinguished from
the critical point, unlike the case $\sigma = 1$, where it is expected to
occur at a coupling $K < K_c$.

\subsection{The three-state Potts chain}
The ferromagnetic Potts model provides a particular generalization of the Ising
model with respect to the number~$q$ of possible coexisting ordered phases. For
$q = 2$ the Ising model is recovered; for $q > 2$ the ferromagnetic Potts model
defines a genuine universality class distinct from the Ising and the more
general ${\rm O}(n)$ universality class. The Potts model is of particular
theoretical interest, because the phase transition it describes may be of first
or second order depending on $q$ and the spatial dimension $d$, even in the
absence of symmetry-breaking fields. For nearest-neighbor interactions in $d =
2$ many properties of the Potts model are exactly known \cite{wu82}. In
particular, if the model is in the first-order regime and $q$ is sufficiently
large, the asymptotic finite-size properties of the nearest-neighbor
ferromagnetic Potts model have been established in a rigorous
fashion~\cite{borgs}. For long-range interactions, however, much less is known
and currently available numerical data are limited to rather small
systems~\cite{glumac98}. We demonstrate in the following, that also for the
Potts model the cluster algorithm introduced in Ref.~\cite{lr-alg} can be
combined with the FFT, allowing the numerical treatment of much larger systems.

Again, we concentrate on the case of algebraically decaying interactions $J(x)
\equiv |x|^{-1-\sigma}$. The Hamiltonian of the ferromagnetic Potts chain with
periodic boundary conditions can then be written in the same form as
Eq.~(\ref{eq:1dhamil}), where the Potts spins ${\bf S}(x)$ are unit vectors
which mark the corners of a (hyper)tetrahedron in $q-1$ dimensions. For the
present case $q = 3$ we employ the complex notation
\begin{equation}
\label{eq:sq3}
 {\bf S}(x) \rightarrow {\cal S}(x) \in
  \left\{1, e^{2\pi i/3}, e^{4\pi i/3} \right\} \; ,
\end{equation}
i.e., ${\bf S}(x) \cdot {\bf S}(y) = \Re [{\cal S}(x) {\cal S}(y)^*]$, where
the asterisk denotes the complex conjugate. The spin representation of the
Potts model given by Eq.~(\ref{eq:1dhamil}) is equivalent to the standard
Kronecker representation, but it has the advantage that the configurational
energy is directly accessible by means of the FFT\@. According to mean-field
theory the ferromagnetic Potts model should always show a first-order phase
transition for $q > 2$. For our case of $d = 1$ and algebraically decaying
interactions, one therefore expects the mean-field prediction to be correct for
sufficiently small values $\sigma > 0$ of the decay exponent of the
interaction, i.e., there should be a critical value~$\sigma_c$ separating
first- and second-order behavior.  Mean-field theory provides an important
guideline for the interpretation of our Monte Carlo data, so we briefly
summarize the basic mean-field predictions.  Following Ref.~\cite{wu82} we
introduce the probability $p_\kappa(x)$ that lattice site $x$ is occupied by
the Potts state $\kappa$, $1 \leq \kappa \leq q$, and we define a homogeneous
scalar order parameter $s$ indicating a broken symmetry with respect to Potts
state $\kappa = 1$:
\begin{equation}
\label{eq:mk}
p_1(x) \equiv m_1 = \frac{1 + (q-1)s}{q} \;, \quad
p_\kappa(x) \equiv m_\kappa = \frac{1-s}{q} \;,\; 2 \leq \kappa \leq q \;.
\end{equation}
For a given value of $s$ the mean-field free-energy density $f_{\rm MF}(s)$
(in units of $k_{\rm B}T$) is then obtained as
\begin{equation}
\label{eq:fMFs}
f_{\rm MF}(s) = -K \zeta(1+\sigma) s^2 + \left\{ [1+(q-1)s] \log[1+(q-1)s]
 + (q-1) (1-s) \log(1-s) \right\} / q \; ,
\end{equation}
where $K = J/(k_{\rm B}T)$ denotes the reduced coupling and $\zeta(\alpha)$ is
the Riemann zeta function. Note that the replacement $K \to (q-1)q^{-1} K$
transforms Eq.~(\ref{eq:fMFs}) from the spin representation into the Kronecker
representation. The transition point $K = K_{\rm MF}^t$ from the disordered
phase $s = 0$ to the ordered $(\kappa = 1)$ phase $s = s_{\rm MF}^t$ follows
from standard mean-field arguments~\cite{wu82}:
\begin{equation}
\label{eq:KsMF}
K_{\rm MF}^t \zeta(1+\sigma) = {(q-1)^2 \over q(q-2)} \log(q-1)\; , \quad
s_{\rm MF}^t = {q-2 \over q-1} \;.
\end{equation}
According to Eqs.\ (\ref{eq:mk}) and~(\ref{eq:KsMF}) the distribution function
$P(m_1)$ for a {\em finite\/} system displays three maxima near the transition
temperature $T_{\rm MF}^t \equiv J/(k_{\rm B}K_{\rm MF}^t)$: one at $m_1 = 1/q$
for the disordered phase, one at $m_1 = (q-1)/q$ for the ordered $(\kappa = 1)$
phase, and one at $m_1 = 1/[q(q-1)]$ for the ordered phases with respect to
the remaining Potts states $(\kappa \geq 2)$. Note that all ordered phases
appear with equal probability in the course of the simulation. For $\sigma
\leq 0.4$ and in our case of $q = 3$ these three peaks in $P(m_1)$ are indeed
located very close to their mean-field positions. For $\sigma = 0.6$ the peaks
are still clearly separated, but they occur at positions shifted with respect
to the mean-field predictions and for $\sigma \geq 0.7$ the peaks start to
overlap strongly and can only be identified for very large systems (see below).

Although the algorithm introduced in Ref.~\cite{lr-alg} is by far the most
efficient one for the simulation with spin systems with long-range
interactions, it is not able to deal with first-order phase transitions beyond
a certain system size. The reason is that like the Metropolis algorithm the
Wolff cluster algorithm encounters an activation barrier between states with
and without long-range order, which is set by the energy-density gap between
the disordered and the ordered phase. For a given size of the gap the tunneling
time between disordered and ordered phases, and therefore the required sampling
time, increases exponentially with the system size so that the attainable
system size $N$ is severely limited. This tunneling problem can be solved by
employing the well-established ideas of multicanonical
sampling~\cite{multican}, however, the generalization of the cluster
algorithm~\cite{lr-alg} to an efficient multicanonical algorithm is beyond the
scope of the present article.

\begin{figure}
\centering\leavevmode
\epsfysize \figurewidth
\rotate[r]{\epsfbox{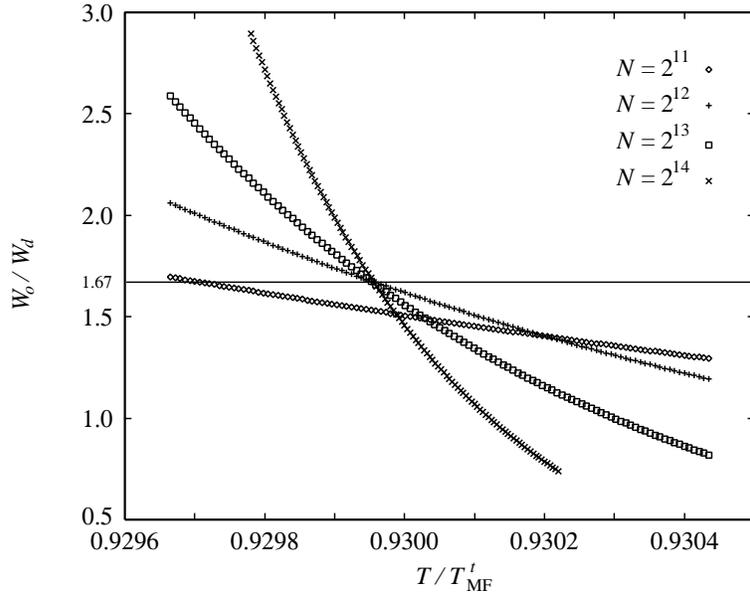}}
\caption[]{Ratio $W_o/W_d$ of the weights of the ordered and the disordered
phase as a function of temperature for the system sizes $N = 2^{11}$
($\diamond$), $2^{12}$ (+), $2^{13}$ ($\Box$), and $2^{14}$ ($\times$).  The
temperature is measured in units of the mean-field transition temperature
$T_{\rm MF}^t$. Statistical errors (one standard deviation, not shown) are
smaller than the symbol sizes. The horizontal line marks the value of $W_o/W_d$
at the intersection of the three largest systems (see main text).}
\label{fig:WoWd}
\end{figure}

\begin{figure}
\centering
\epsfysize \figurewidth
\leavevmode
\rotate[r]{\epsfbox{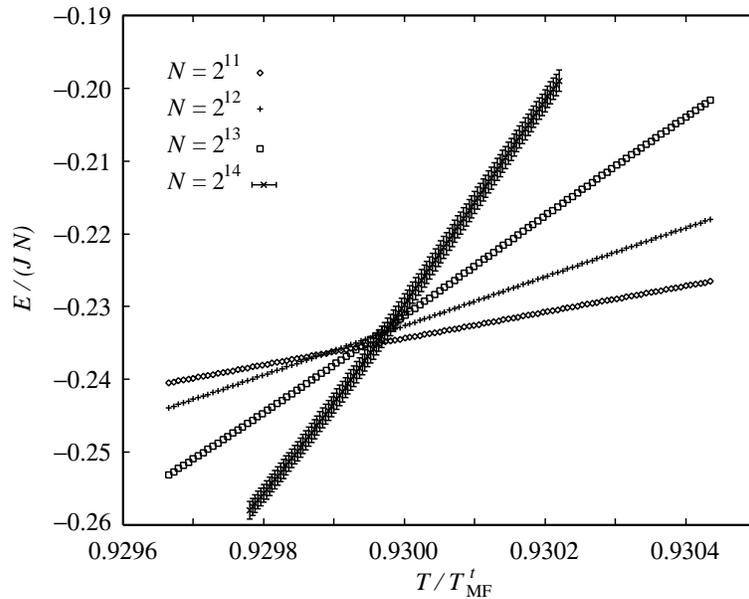}}
\caption[]{Energy $E/(J N)$ per spin in units of the coupling constant $J$
as a function of temperature for the system sizes $N = 2^{11}$ ($\diamond$),
$2^{12}$ (+), $2^{13}$ ($\Box$), and $2^{14}$ ($\times$). The temperature is
measured in units of the mean-field transition temperature $T_{\rm MF}^t$.
Statistical errors (one standard deviation) are only shown when they exceed the
symbol sizes. Within the statistical error the curves for the three largest
systems intersect at the same temperature as $W_o/W_d$ in
Fig.~\protect\ref{fig:WoWd}.}
\label{fig:energy}
\end{figure}

\begin{figure}
\centering\leavevmode
\epsfysize \figurewidth
\rotate[r]{\epsfbox{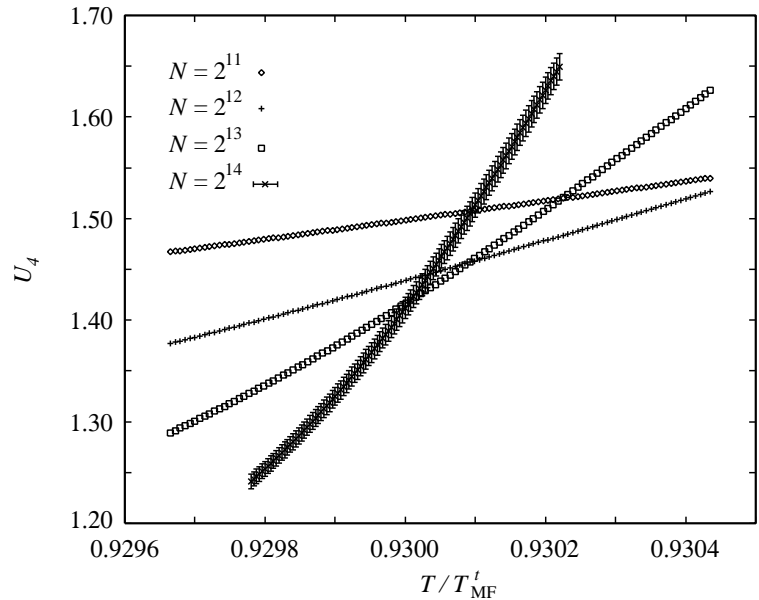}}
\caption[]{Fourth-order energy cumulant $U_4$ as a function of temperature for
the system sizes $N = 2^{11}$ ($\diamond$), $2^{12}$ (+), $2^{13}$ ($\Box$),
and $2^{14}$ ($\times$). The temperature is measured in units of the mean-field
transition temperature $T_{\rm MF}^t$.  Statistical errors (one standard
deviation) are only shown when they exceed the symbol sizes. The curves do not
have a common intersection within the displayed temperature range indicating
much larger finite-size effects than in Figs.\ \protect\ref{fig:WoWd}
and~\protect\ref{fig:energy}.}
\label{fig:U4}
\end{figure}

The data we present in the following have been obtained from histograms of the
energy taken at several temperatures. The data are again conveniently analyzed
by the optimized multiple-histogram method~\cite{multiple-hist}.  For the
values $\sigma = 0.2$ and $\sigma = 0.4$ the Potts chain undergoes a strong
first-order phase transition~\cite{glumac98} which limits the chain length to
$N = 2^{13}$ in the former case and to $N = 2^{14}$ in the latter case. We
reinvestigate chains from $N = 2^{10}$ spins to the respective maximum chain
length by taking a few times $10^6$ independent samples for
each system size and temperature, where a comparison with the finite-size
theory of Borgs {\em et al.}~\cite{borgs} turns out to be very instructive. We
restrict the detailed presentation to the case $\sigma = 0.4$; the data for
$\sigma = 0.2$ are qualitatively very similar. Near the transition temperature
the energy distribution function~$P(E)$ displays two peaks characterizing the
ordered and the disordered phase, respectively \cite{2dPotts}. In
Ref.~\cite{glumac98} the temperature of equal peak height is taken as an
estimate for the transition temperature on the finite system, where the leading
finite-size corrections are of the order ${\cal O}(1/N)$.  If the systems are
large enough, i.e., the peaks in the energy distribution are well separated,
the ratio $W_o/W_d$ of the weight of the ordered phase~$W_o$ and the weight of
the disordered phase~$W_d$ provides a far more convenient indicator of the
transition temperature, because the associated finite-size corrections decay
{\em exponentially\/} with $N$ \cite{borgs,2dPotts}. Our result for long-range
interactions is shown in Fig.~\ref{fig:WoWd}. The transition temperature is
marked by the intersection of $W_o/W_d$ as a function of temperature for the
three largest systems. For $N = 2^{11}$ the peaks in the energy distribution
are not well separated so that $W_o/W_d$ is not well defined in this case. For
$N \geq 2^{12}$ the curves meet at $W_o/W_d = 1.67~(2)$ as shown by the solid
line. For $\sigma = 0.2$ we find a corresponding intersection at $W_o/W_d =
1.25~(2)$. For nearest-neighbor interactions in $d \geq 2$ and sufficiently
large~$q$ the value $W_o/W_d = q$ is expected to indicate the transition
temperature~\cite{borgs,2dPotts}.  Surprisingly, we find a much smaller value
here which appears to increase with~$\sigma$. From Ref.~\cite{borgs} one
furthermore expects that the curves displaying the energy density for different
system sizes as a function of temperature exhibit an intersection close to the
transition temperature, where the deviations are predicted to be exponentially
small in $N$. In Fig.~\ref{fig:energy} this situation is shown for long-range
interactions with $\sigma = 0.4$. A corresponding result has been obtained for
$\sigma = 0.2$.  The energy densities intersect near the transition temperature
found in Fig.~\ref{fig:WoWd}, where the shifts between mutual intersections
seem to be compatible with exponentially small finite-size effects.  Still too
few data are available for a quantitative analysis of these shifts, but
finite-size effects of the order $1/N$ can be ruled out. The fourth-order
energy cumulant $U_4$ defined by
\begin{equation}
\label{eq:U4}
U_4 \equiv \langle {\cal H}^4 \rangle / \langle {\cal H}^2 \rangle^2
\end{equation}
is shown in Fig.~\ref{fig:U4} for different system sizes as a function of
temperature, where $\cal H$ is the Hamiltonian given by Eq.~(\ref{eq:1dhamil}).
These cumulants should also intersect at the transition temperature in the
limit $N \to \infty$. The data displayed in Fig.~\ref{fig:U4} show the expected
tendency, but the finite-size corrections are much larger than those for the
weight ratio or the energy density (see Figs.\ \ref{fig:WoWd}
and~\ref{fig:energy}, respectively). For $\sigma = 0.2$ a corresponding result
has been found. The systematic shift of the intersections of $U_4$ for
different system sizes is compatible with a $1/N$ behavior as anticipated in
Ref.~\cite{borgs} for nearest-neighbor interactions, but the present amount of
data is too limited to give reliable quantitative evidence for this behavior.

\begin{figure}
\centering\leavevmode
\epsfysize \figurewidth
\rotate[r]{\epsfbox{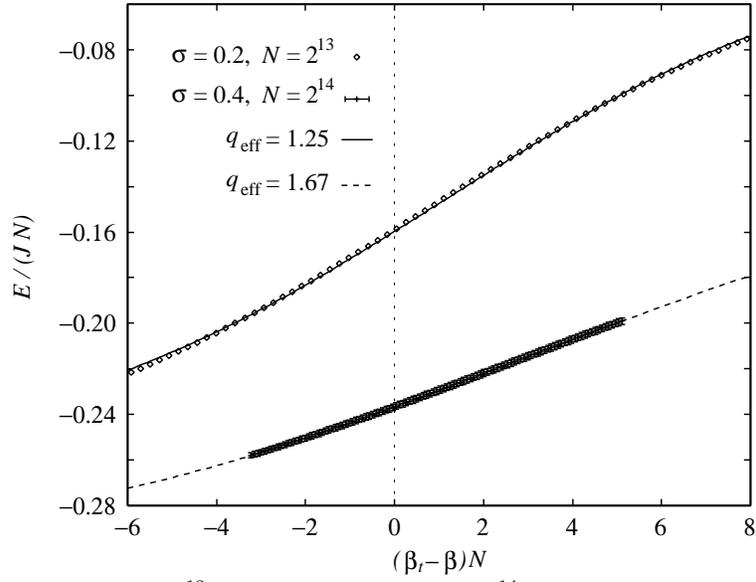}}
\caption[]{Energy density for $\sigma = 0.2$, $N = 2^{13}$ ($\diamond$) and
for $\sigma = 0.4$, $N = 2^{14}$ (+) as functions of the scaling variable
$(\beta_t - \beta)N$ in comparison with Eq.~(\protect\ref{eq:ebetaL}) for $q
\to q_{\rm eff}(\sigma=0.2) = 1.25$ (solid line) and $q \to q_{\rm
eff}(\sigma=0.4) = 1.67$ (dashed line). Error bars on the numerical data (one
standard deviation) are only shown when they exceed the symbol sizes. The
inverse temperature $\beta = 1/(k_{\rm B}T)$ is given in units of $1/(k_{\rm B}
T^t_{\rm MF})$.}
\label{fig:e2e4}
\end{figure}

For nearest-neighbor interactions and periodic boundary conditions the energy
density asymptotically obeys the scaling law (cf.\ Eq.~(1) of
Ref.~\cite{borgs})
\begin{equation}
\label{eq:ebetaL}
E(\beta,L) \simeq {E_d + E_o \over 2} - {E_d - E_o \over 2} \tanh \left[
{E_d - E_o \over 2} (\beta - \beta_t) N + {\log q \over 2} \right],
\end{equation}
where $\beta = 1/(k_{\rm B}T)$ is the inverse temperature and $\beta_t$ is the
transition point. It is instructive to compare the scaling form given by
Eq.~(\ref{eq:ebetaL}) with the data displayed in Fig.~\ref{fig:energy}.  The
energies of the disordered phase $E_d$ and the ordered phase $E_o$ can be read
off from the positions of the two maxima of the energy distribution function.
It turns out that the data in Fig.~\ref{fig:energy} and their counterpart for
$\sigma = 0.2$ are in fact consistent with Eq.~(\ref{eq:ebetaL}) within the
error bars, {\em provided\/} the number of states~$q$ on the right-hand side of
Eq.~(\ref{eq:ebetaL}) is replaced by the {\em effective\/} value $q_{\rm
eff}(\sigma = 0.4) \equiv W_o/W_d|_{\beta=\beta_t} = 1.67$ measured in
Fig.~\ref{fig:WoWd} at the transition temperature or its counterpart $q_{\rm
eff}(\sigma = 0.2) = 1.25$, respectively. The quantitative
comparison of our data for $\sigma = 0.2$, $N = 2^{13}$ and $\sigma = 0.4$, $N
= 2^{14}$ with Eq.~(\ref{eq:ebetaL}) is shown in Fig.~\ref{fig:e2e4}. Within
the statistical errors, the agreement is excellent except for larger values of
the scaling variable $(\beta - \beta_t) N$. These deviations are due to the
fact that Eq.~(\ref{eq:ebetaL}) only holds asymptotically for sufficiently
large systems. For finite systems additional finite-size corrections enter
through the residual $N$-dependence of $E_d$, $E_o$, and $q_{\rm eff}(\sigma)$
which appear as parameters in Eq.~(\ref{eq:ebetaL}).  Figure~\ref{fig:e2e4}
demonstrates that the finite-size effects in the three-state Potts chain with
periodic boundary conditions and long-range interactions can be interpreted in
terms of the Borgs--Koteck\'y theory \cite{borgs} for the nearest-neighbor
Potts model in higher dimensions for an effective number of states $q_{\rm
eff}(\sigma)$. The physical meaning of $q_{\rm eff}(\sigma)$, however, remains
unclear. The proof of Eq.~(\ref{eq:ebetaL}) also requires the assumption that
$q$ is sufficiently large~\cite{borgs}, so $q = 3$ may not be sufficient for a
quantitative comparison. On the other hand, numerical investigations have shown
that the $q \geq 5$ nearest-neighbor Potts model in $d = 2$~\cite{2dPotts} and
the $q = 3$ nearest-neighbor Potts model in $d = 3$~\cite{janke} follow the
theory of Ref.~\cite{borgs} very closely despite the small values of~$q$.
Further analytical and numerical studies are required to settle this question.

\begin{figure}
\centering\leavevmode
\epsfysize \figurewidth
\rotate[r]{\epsfbox{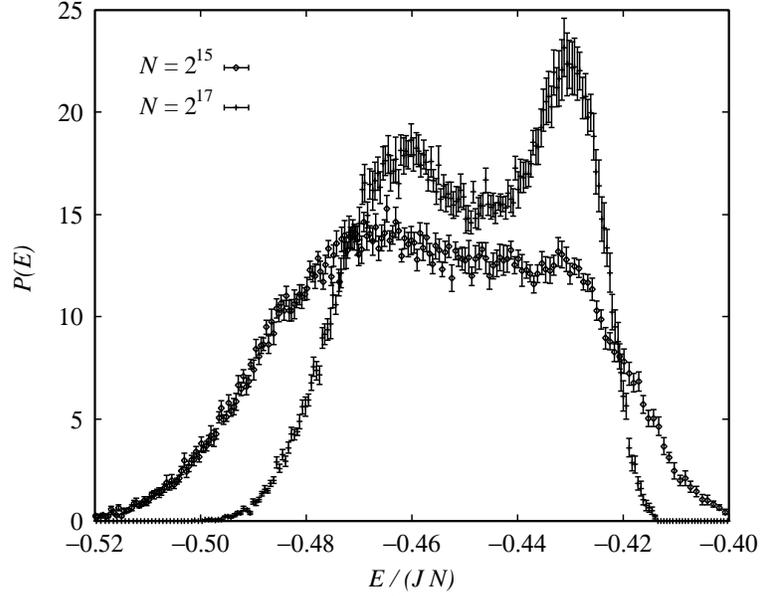}}
\caption[]{Energy distribution function $P(E)$ for $\sigma = 0.7$, $T/T^t_{\rm
MF} = 0.8095$, and system sizes $N = 2^{15}$ ($\diamond$) and $N = 2^{17}$ (+).
The typical double-peak structure remains invisible for $N < 2^{16}$ spins but
sharpens if $N$ is increased at fixed $T$.}
\label{fig:PE7}
\end{figure}

We close our discussion of the three-state Potts chain with a brief summary of
our results for $\sigma = 0.6$, 0.7, and 0.75, which have been studied with
reduced statistics ($10^5$ independent samples for each system size and
temperature).  The value $\sigma = 0.6$ is still located in the first-order
regime~\cite{glumac98}, but in order to obtain a well-defined weight ratio
$W_o/W_d$ system sizes of $N \geq 2^{16}$ Potts spins are required, although
the maxima in the energy distribution are well separated already for $N \geq
2^{14}$. We have performed simulations for $N = 2^{14}$ up to $N = 2^{17}$ at
four to six temperatures for each system size.  Even for $N = 2^{17}$ the
finite-size corrections are too large to identify an intersection of the energy
densities as accurately as displayed in Fig.~\ref{fig:energy}. Data acquisition
for $N > 2^{17}$ is strongly hampered by the energy gap so that we refrain from
discussing our data for $\sigma = 0.6$ in any more detail. In contrast to
Ref.~\cite{glumac98} the value $\sigma = 0.7$ of the decay exponent can
undoubtedly be identified as a member of the first-order regime. For system
sizes $N \geq 2^{16}$ the energy distribution function displays the typical
double-peak structure which becomes sharper as the system size is increased at
fixed temperature. We illustrate this in Fig.~\ref{fig:PE7}, where the data for
$P(E)$ are shown at $T/T^t_{\rm MF} = 0.8095$, which is close to the transition
point. The same analysis has been repeated for $\sigma = 0.75$ and system sizes
up to $N = 2^{19}$ spins. Although $P(E)$ also develops a plateau similar to
the one displayed in Fig.~\ref{fig:PE7} for $N = 2^{15}$, no double-peak
structure could be resolved up to $N = 2^{19}$ so that $\sigma = 0.75$ may
already belong to the second-order regime of the three-state Potts chain with
long-range interactions. However, the detection of a double-peak structure in
$P(E)$ for a given value of $\sigma$ is essentially a matter of attainable
system size (see Fig.~\ref{fig:PE7}), so $\sigma_c > 0.7$ is the only safe
conclusion here.

\section{Conclusions}
The combination of the recently developed cluster algorithm~\cite{lr-alg} for
systems with long-range interactions with the Fast Fourier Transform for the
calculation of the configurational energy leads to a Monte-Carlo algorithm with
a very high efficiency. In particular, the FFT allows to extend the attainable
system sizes by two orders of magnitude in comparison with other approaches
(cf.\ Ref.~\cite{glumac98}). Histogram interpolation methods then allow the
investigation of thermodynamic properties of these systems with unprecedented
resolution. By construction the algorithm can only deal with first-order phase
transitions up to a limited system size. In order to avoid this limitation the
algorithm must be generalized to include multicanonical sampling. Here, we have
demonstrated the potential of the algorithm for the Ising chain and the
three-state Potts chain with algebraically decaying interactions. For
completeness, it is mentioned that also for system sizes that are not integer
powers of two a considerable gain can be obtained by performing the discrete
Fourier Transform via, e.g., a prime-factor algorithm.

For the Ising chain we have investigated the finite-size behavior of the
specific heat in the classical regime for $\sigma = 0.25$ and in the
nonclassical regime for $\sigma = 0.9$. In the former case the specific
heat behaves essentially mean-field like, i.e., the expected discontinuity
in the specific heat at the critical temperature in the thermodynamic limit
builds up as the system size is increased. On the other hand, the choice
$\sigma = 0.9$ is expected to yield a negative specific-heat exponent, i.e.,
a cusp singularity should appear with increasing system size. Our numerical
data confirm this behavior as well and clearly show the different shapes of the
specific-heat curves in the two cases.

The three-state Potts chain is expected to show a first-order phase
transition for $\sigma < \sigma_c$, where our results indicate that
$\sigma_c > 0.7$. For $\sigma = 0.2$ and $\sigma = 0.4$, for which the
$q = 3$ Potts chain displays a strong first-order phase transition, our
data confirm the Borgs--Koteck\'y scenario of the first-order transition
in Potts models with nearest-neighbor interactions in higher dimensions,
provided the number $q$ of states is replaced by the {\em effective\/} number
of states $q_{\rm eff}(\sigma) = W_o/W_d|_{\beta=\beta_t} < q$ which also
enters the finite-size scaling form of the energy density near the
transition temperature. For $\sigma = 0.6$ the same behavior can be
confirmed only on a semi-quantitative level, because much larger systems
must be investigated in order to obtain sufficient resolution. The mechanism
that leads to the reduction of the effective number of states and the
physical interpretation of $q_{\rm eff}(\sigma)$ are not known.

\acknowledgments
We gratefully acknowledge helpful discussions with T.~Neuhaus and W.~Janke, and
stimulating comments by K.~Binder. Furthermore, the authors wish to thank the
organizers of the Twelfth Annual Workshop on Recent Developments in Computer
Simulation Studies (Athens, Georgia), where the collaboration leading to this
work was initiated.  M.~Krech also gratefully acknowledges financial support
through the Heisenberg program of the Deutsche Forschungsgemeinschaft.


\newpage

\end{document}